# New Zealand's size and isolation can stimulate new science

Philip Yock[1]

*Abstract: It is conventionally thought that New Zealand's distance from the large, northern hemisphere centres of learning, and our relatively small population and wealth, are detrimental to the contribution we can make to the advancement of scientific knowledge. Here the reverse point of view is argued.*

## Introduction

Dame Anne Salmond recently described New Zealand as an increasingly exciting and urbane country that celebrates diversity and reaches out to the world (Salmond 2015). She exhorted New Zealanders to set their sights high to make life in New Zealand the best it can be – inventive, entrepreneurial, exciting and generous in spirit.

Whilst one would wish to agree warmly with Dame Anne's sentiments, it has been my experience that there is opposition to such sentiments in some scientific circles at the University of Auckland where the accepted wisdom appears to have been that New Zealand is too small and too isolated to initiate new developments in science.

This is the impression I gained doing physics at the University of Auckland over a period of several years, and it is not one that I should like to leave the field with. It is my belief that New Zealand's size and isolation can stimulate new science. This article is penned to promote this point of view.

In what follows two specific examples from physics are described where new developments were pioneered locally with some measure of success, but which were met with scepticism that continues to the present day.

In one of the examples to be described below a search for planets orbiting the stars of the night sky was commenced before any such planets were known to exist, and in the other a precursor to today's Standard Model of matter was formulated. These were the final and initial fields of science that the author contributed to.

Before recounting these examples it is noted that Dame Anne was not alone with her comments referred to above. A few months ago Professor Kathleen Campbell of the University of Auckland and I urged emerging Kiwi scientists and aspiring school leavers to set their sights high in their choices of fields to study, and to consider fields that specifically make use of New Zealand's geographical and environmental context following inspiring examples set by our sportspeople, writers and others (Campbell & Yock 2015). Needless to say, I stand by these comments. What could be more inspiring than our west coast with flax, gannet colonies, surf, stars, clouds and Aurora Australis at Muriwai as depicted below in Figure 1?

[1] 2/42 Reihana Street, Orakei, Auckland, NZ
p.yock@auckland.ac.nz

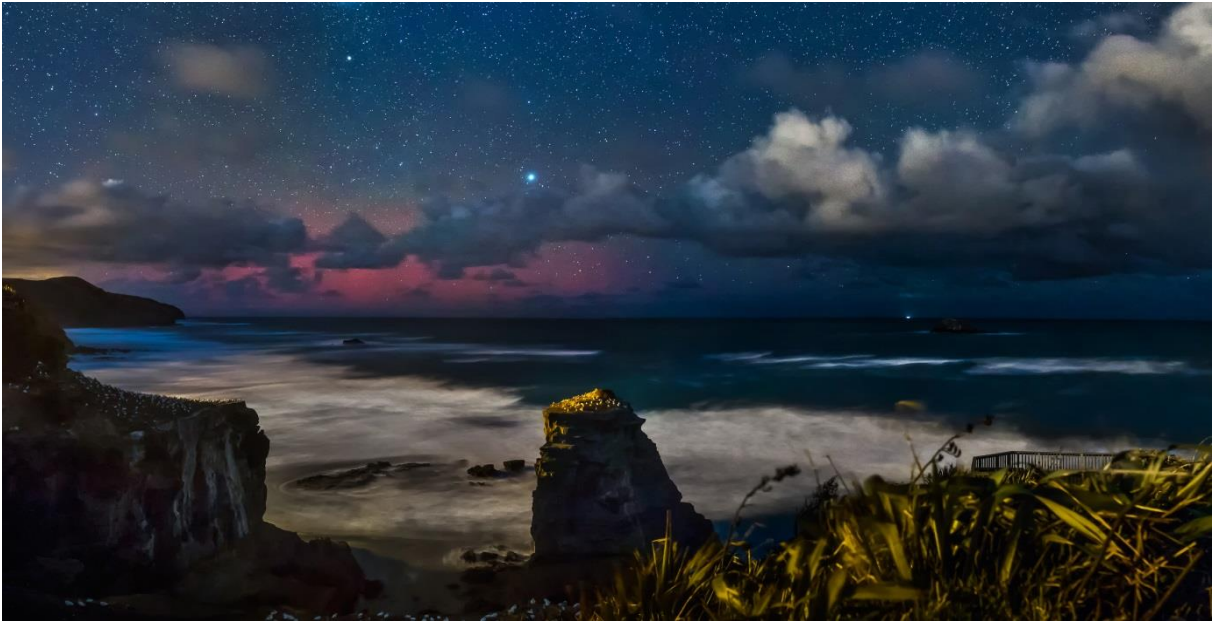

*Figure 1. New Zealand coastline at Muriwai. Recent observations, including some made from New Zealand which are described below, have shown that approximately one in five stars like those shown above host a habitable planet like Earth, and one in two an ice-giant planet like Neptune. Photograph courtesy Jonathan Green and Amit Ashok Kamble.*

At about the same time that Campbell and I wrote the above paper, Professor Sir Peter Gluckman, the NZ Science Adviser to the Prime Minister wrote "*while New Zealand is a modest component of the international research effort … our contribution to the global effort in discovery science should be protected*" (Gluckman 2015a). And in another recent publication he wrote "*science communication is an inherent part of the scientific enterprise, and it needs integrity if the reliability of science is to be protected*" (Gluckman 2015b). I concur with these statements.

## Planets orbiting the stars of the Milky Way

The stars of the night sky have long awakened the interest of philosophers and scientists to the possible presence of civilizations out there inhabiting planets orbiting the stars. Giordano Bruno, Isaac Newton and Alexander Pope were amongst the early proponents of the idea, and possibly most of us have had related thoughts from time to time, especially in our youths.

The vast distances to the stars, however, prevented tangible progress being made until recent times. It is in fact quite difficult to appreciate the immense distances that are involved. The late UK physicist Sir James Jeans put it well when he said *"put three grains of sand in a vast cathedral, and the cathedral will be more closely packed with sand than space is with stars" (*Jeans 1987)*.*

The search for extra-terrestrial intelligence via radio signals was pioneered last century soon after communication by radio was first demonstrated, although the sensitivity required to eavesdrop on intra-planetary communications on extra-solar planets[2] had certainly not been reached then. The next generation of radio telescopes, beginning with the 500m FAST telescope in China, may be the first to reach this milestone (Nan et al. 2011). The FAST telescope is due to commence operations

---

[2] "Extra-solar planet" is the term used for a planet orbiting a star like the Sun. Nowadays it is commonly shortened to "exoplanet".

this year (Wong 2016).

In addition to this US$100M was recently donated by the Russian billionaire Yuri Milner for the purpose of seeking signs of extra-terrestrial intelligence in a 10-year project dubbed *"Breakthrough Listen"* that will use the Green Bank and Parkes radio telescopes in USA and Australia, and the Lick Observatory in California (Merali 2015).

Probably the majority viewpoint is that these efforts will fail if only because of the finite time span over which advanced civilizations are likely to survive. Two civilizations may need to be relatively nearby in both space and time to make contact, and this will reduce the odds of success in the above projects. The chances of testing our theories here on Earth by comparison with those of more advanced civilizations may be slim.

In the meantime successful advances have been made on the less ambitious task of detecting planets orbiting stars in the Milky Way, including terrestrial planets like Earth. This was not an easy task, as small planets like Earth orbiting stars like the Sun cannot be seen directly with existing telescopes. The planets are too close to their parent stars to be resolved, and they are lost in the glare of their parent stars. This necessitated the use of indirect detection techniques.

Several nations contributed to the hunt for extra-solar planets orbiting Sun-like stars following the first detection that was made by Swiss astronomers in 1995 (Mayor & Queloz, 1995). The search for, and the study of, these exoplanets is now the fastest growing field in astronomy.

Several techniques are in use and their sensitivities have gradually been honed to the stage where Earth-like planets are just becoming detectable. In 2013 a US team estimated that approximately 22% of Sun-like stars host an Earth-like planet in the habitable zone where liquid water could exist (Petigura et al. 2013).

The astronomical community is presently awaiting confirmed signals of such planets orbiting our neighbour stars Alpha Centauri A and B. These are near the Southern Cross and a search for exoplanets orbiting them is presently underway from the Mt John Observatory in Canterbury (Yock 2015).

*Gravitational microlensing*
NZ astronomers have specialized in an exotic technique known as "gravitational microlensing" to hunt for exoplanets. Gravitational microlensing is a modern term that refers to the bending of light by the gravitational field of a star.

In 1936 Einstein predicted that the gravitational field of a star could act like a large lens, and magnify a more distant star, but he did not expect the effect to have practical applications (Einstein 1936). Recent studies with modern computerized telescopes have however exceeded Einstein's expectations. Gravitational microlensing has been found to provide a surprisingly sensitive means for detecting planets orbiting stars in our galaxy, the Milky Way.

The lensing effect occurs when two distant stars are very well aligned when observed from Earth. Such alignments are most likely to occur in the dense star fields towards the centre of the galaxy which is in the southern constellation Sagittarius. New Zealand is therefore well placed to observe the effect, and, since 2004, the world's largest telescope dedicated to gravitational microlensing has been the 1.8m Japan/NZ telescope known as MOA which is located at the Mt John University Observatory in Canterbury (Yock 2006; Yock 2012; Yock 2015).

Planets found by microlensing generally have orbital radii about their host stars that are about three times larger than the orbital radii of habitable planets like Earth. They occupy a region of space that other planet detection techniques are not suited to. Microlensing planets are too cold for life like ours to have evolved on them, but nevertheless their mere discovery provides useful information on the types of planets that are present in the galaxy, and on the physical processes that resulted in their formation.

Astronomers from several countries, including New Zealand, have now detected about 50 cool planets by Einstein's lensing effect. Beginning this year the discovery rate is expected to increase to about 50 per year thanks to new telescopes being deployed by Korea (Henderson et al. 2014). And in 2024 another large jump will occur with the launch by NASA of a dedicated space telescope known as WFIRST (Gaudi 2016). In addition, the Kepler Space Telescope will be used in a special campaign this year to observe the lensing effect simultaneously from the ground and from space to enable triangulation measurements to be conducted (Henderson et al. 2016). New Zealanders are involved in all these projects.

It has already been reported that approximately 52% of all stars in the Milky Way possess planets similar to Neptune (Cassan 2012). These ice-giant planets are likely to have formed by the absorption of ices on terrestrial embryos beyond the "snowlines" of young stars (Ida & Lin, 2004). Besides being the major source of material for Neptune-like planets, these ices are also likely to be the major source of water on warmer, terrestrial planets like Earth.

New Zealand contributed positively to all the above. Unfortunately, however, the discovery of the fourth planet by Einstein's lensing process broke the trend. This discovery was questioned at the University of Auckland on grounds that were neither scientifically sound nor factually correct. This is described below in the hope that future discoveries will be reported correctly.

*The fourth planet found by microlensing*
Prior to the discovery of the fourth microlensing planet it had been noted at the University of Auckland, and elsewhere, that Einstein's lensing effect could produce high magnifications, of order 100 or more, and that these magnifications, although rare, could result in high sensitivity to the presence of planets. This was not popular thinking at the time, but the idea was nevertheless promoted enthusiastically at the University of Auckland (Rattenbury et al. 2002). I recall our enthusiasm being driven in part by our isolation from other groups with access to larger telescopes. History has shown that our enthusiasm was not misplaced. The majority of planets discovered by microlensing to the present time were found by the high magnification technique (Shvartzvald et al. 2016).

The lensing event in which the fourth microlensing planet was found occurred in 2005. It reached a magnification of 800 and it was therefore sensitive to the presence of planets. Visual inspection of the data showed possible evidence of a planet, and I made plans for a group I was then leading at the University to fully analyse the data using computer code that had just been written at the University (Yock 2015). A US group followed suit and proceeded to analyse the same data independently.

Consistent results were obtained in the NZ and the US analyses and these were published jointly in the normal manner (Gould et al. 2006). This publication was the first to report a high abundance of ice-giant planets like Neptune.

Recently a new test of the measurements became possible using the Hubble Space Telescope and the Keck Observatory in Hawaii. This was duly carried out and successful results obtained. These were published in 2015 in the normal manner (Bennett et al. 2015; Batista et al. 2015). All the above was described recently in a non-specialist article (Yock 2015).

On the basis of yet further studies conducted with several former students of the University (Abe et al. 2013) I am hopeful that future observations at high magnification will reveal that stars commonly host pairs and possibly even triplets of Neptune-like planets (Yock 2016). Intriguingly, at the time of writing, a possible new member of our Solar System has just been announced which may be yet another Neptune-like planet (Witze 2016). Also, at the time of writing, a new Neptune-like planet has just been found using the techniques proposed by Abe et al in the above paper (Koshimoto et al. 2016). Neptune-like planets are turning up frequently and I believe they will continue to do so.

***Reaction to the discovery of the fourth microlensing planet at the University of Auckland***
Unfortunately the validity of the results that were obtained at the University of Auckland on the fourth microlensing planet, and also the techniques that were used to acquire them along with my scientific integrity, were strenuously questioned by staff of the University (Walls 2012). These staff turned a blind eye to the results that had been obtained by my group at the University and claimed that a member of the group had been allocated inadequate computing resources. They authorised a year-long re-analysis of the data with 100 computers using code that was known to contain errors. This rather defied common-sense and, not surprisingly, yielded results that were not useful.

Normal procedures had in fact been followed in the analysis carried out at the University, and those who were critical had either failed to acquaint themselves with these procedures, or misconstrued the results, or both. To address these issues the procedures that had been used at the University, and also the results obtained, were published recently (Yock 2015). Both were consistent with those reported in the original publication (Gould et al. 2006).

Also, as noted above, new and independent observations were made recently with the Hubble Space Telescope and with the Keck Observatory that confirmed the original results (Bennett et al. 2015; Batista et al. 2015). The number of independent confirmations of the original results has thus now risen to three. However, on being informed recently of this, University management declined to reconsider their assessment of the work. This action, or lack of it, appeared to fly in
the face of the recommendations of the Prime Minister's Science Advisor for protecting discovery science in New Zealand and reporting research correctly.

Thus concluded a tragedy of errors in the history of astronomy in which the scientific method was abandoned and the science unjustifiably misrepresented for no good purpose. Further discussion on this appears below.

## The nature of the strong nuclear force
Matter is composed of atoms which may be likened to tiny planetary systems orbiting stars. The orbiting bodies in atoms are of course electrons and they orbit an atomic nucleus that is composed of neutrons and protons. We have Rutherford to thank for this picture. He deduced this picture from the results of his well-known experiment in which alpha particles from a radio-active source were fired at gold foil.

In what follows, a brief account is given from an unabashedly personal perspective of subsequent work that led to current understanding of the nature of the strong force that holds the atomic nucleus together, including a contribution from New Zealand.

*Experiments from the Rutherford era on the strong nuclear force*
Rutherford's experiment with gold foil did not reveal the nature of the strong nuclear force. Coulomb repulsion between the alpha particle and the gold nucleus was sufficient to ensure they never came into contact.

However, Rutherford and his contemporaries followed up the gold foil experiment with a series of experiments that used lighter targets, including magnesium and aluminium (Rutherford & Chadwick 1925). In these experiments the alpha particle was able to enter the target nucleus and thereby probe the nature of the nuclear force. These experiments were the forerunner of today's experiments conducted at the CERN laboratory in Switzerland in which high-energy protons or nuclei are collided head-on in the Large Hadron Collider.

The experiments from the Rutherford era with light targets revealed the atomic nucleus to be, with high probability, an approximately spherical system of closely packed protons and neutrons that were themselves tiny spheres of diameter approximately $2.6 \times 10^{-15}$ m as depicted in Figure 2 (Pollard 1935; Evans 1955). This historical measurement of the neutron and proton diameters was beautifully confirmed in high-precision electron scattering measurements that were conducted at Stanford University 20 years later (Hahn et al. 1956).

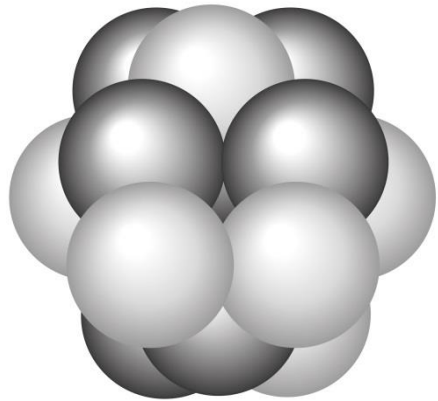

*Figure 2. The atomic nucleus was found in the 1930s to be an approximately spherical assemblage of closely packed protons and neutrons of diameter approximately $2.6 \times 10^{-15}$ m.*

The above result provides significant information on the nuclear force. Neutrons and protons that are separated by more than a nuclear diameter do not experience the nuclear force, although the cut-off is not perfectly sharp. If they approach one another more closely than a nuclear diameter they first feel the attraction of the nuclear force, but eventually this turns into a stronger repulsive force to maintain the constant density of nuclear matter for all elements that was reported by Pollard in 1935 and is implied by the above figure.

It was also found in these early days, notably by Francis Aston at Cambridge University, that the force that acts between neutrons and protons is very strong. Aston weighed the nuclei of various nuclear species using an accurate mass spectrograph (Aston 1936). From the masses so obtained it was possible to deduce the binding energies of the nuclei through use of Einstein's equation $E = mc^2$. The results implied that the binding energies of neutrons and protons in nuclei are some millions of times greater than the binding energies of outer electrons in atoms. In other words, the forces that bind neutrons and protons together in nuclei were found to be very much stronger than the chemical forces that bind atoms together in molecules. This, of course, was the origin of the nuclear age.

Nowadays we understand the strong nuclear force to be one of the four fundamental forces of nature, the others being gravity, electromagnetism and the weak nuclear force. The latter force is responsible for the beta type of radioactivity which is used in PET scans, for example. The weak and the strong nuclear forces are jointly responsible for the production of energy in stars like the Sun. It is of course the latter energy that permitted life to evolve on Earth, and also presumably on other planets orbiting other stars in the Milky Way.

### *New particles discovered from the 1930s to the 1960s*
In the thirty years that followed the Rutherford era many new types of particles were discovered. Most were found in the cosmic radiation[3] and most were unexpected.

Several of these discoveries were made by observing the tracks of cosmic rays in a device known as the "cloud chamber". This revealed the tracks of charged cosmic rays passing through as trails of droplets in super-saturated vapour. By studying the effects of magnetic fields and steel absorbers on the particles it was possible to determine their main properties. A cloud chamber that was built at the University in the Auckland in the 1950s is shown below. It revealed tracks of cosmic rays at a rate of a few per second.

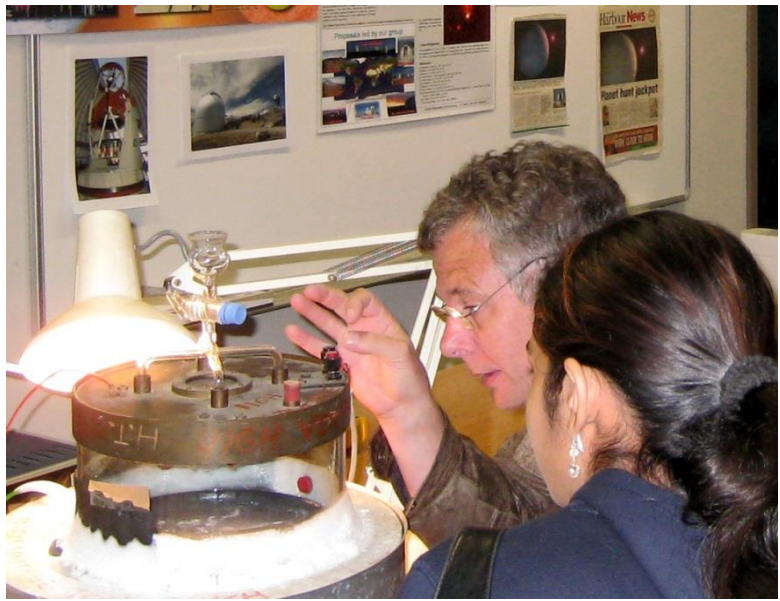

*Figure 3. A cloud chamber from the 1950s being demonstrated at an open day at the University of Auckland by Mr Denis Taylor in about 2005.The tracks of cosmic rays appear in the super-saturated vapour in the chamber.*

---

[3] The cosmic radiation is a flux of high-energy particles comprised mainly of electrons, protons and atomic nuclei that is confined within the Galaxy by its magnetic field. Their origin is unknown although it is possible that most are produced in the remnants of supernova explosions (Yock 2012).

Another popular detector for cosmic rays that was also used at the University of Auckland was the "photographic emulsion". This revealed the tracks of particles as developed silver grains. It had the advantage of being able to be flown with balloons and exposed to the primary cosmic radiation at altitudes up to 30 km near the geomagnetic poles (Rao & Yock 1987).

In the 1950s and 1960s experiments with cosmic rays gradually gave way to experiments with particle accelerators which, although they could not reach such high energies, enabled observations to be made under controlled conditions.

The most prominent discoveries of particles made from the 1930s to the early 1960s are tabulated below. With the exception of the neutrino, all the new particles were produced in collisions in which kinetic energy was converted to mass via Einstein's equation $E=mc^2$. The neutrino was discovered as a product of a nuclear power reactor.

Today's Standard Model of matter was constructed largely on information contained in Table 1. It is therefore worthwhile to expand briefly on the entries.

| Particle | Strongly interacting? | Lifetime (approx) | Mass/$m_e$ | Predicted by | Discovery |
|---|---|---|---|---|---|
| Positron, $e^+$ | No | Infinite | 1 | Dirac, 1928 | Anderson 1933 |
| Muon, $\mu^\pm$ | No | $10^{-6}$ s | 205 | - | Neddermeyer & Anderson 1937 |
| Pion, $\pi^\pm$, $\pi^0$ | Yes | $10^{-8}$, $10^{-16}$ s | 270 | Yukawa, 1935 | Lattes et al. 1947 |
| Strange particles, K, Λ, Σ, Ξ, $\Omega^-$ | Yes | $10^{-10}$ s | 967 – 3272 | - | Butler & Rochester 1947… |
| Neutrino, ν | No | ? | ? | Pauli, 1933 | Reines & Cowan 1956 |
| Resonances, N*, ρ, ω, … | Yes | $10^{-23}$ s | > 1500 | - | Anderson et al. 1952; Erwin et al. 1961 … |

***Table 1. Main properties of particles discovered from the 1930s to the early 1960s. It is often said that the $\Omega^-$ was not discovered until after the naïve quark model (see below) was proposed, but the first sighting was actually made in 1954 using photographic emulsion (Eisenberg 1954).***

The first particle, the positron, represents one of the greatest triumphs of science. This is the anti-electron, the first known example of anti-matter. It has the same mass as the electron but the opposite charge. A positron annihilates on coming into contact with an electron with the emission of radiation, a process that is utilised in PET scans today. The UK theorist Paul Dirac formulated an equation in 1928 which describes the properties of the positron very accurately, although he was hesitant to interpret it in terms of the positron before the experimental discovery of the particle (Dirac 1928). Nowadays it is taken for granted that all charged particles have associated with them antiparticles with the same mass and the opposite charge.

The second particle, the "muon", appeared to be mysterious on its discovery, and has remained so ever since. The reason for its existence was not apparent when it was discovered in 1937, and this situation has not changed since then (Feynman 1985). The muon does not affect the properties of normal matter to an appreciable degree, and the evolution of planets, stars and galaxies would, as far as is known, be unaffected if the muon did not exist. On being told of its discovery in 1937 it is said that the well-known nuclear physicist of the day, Isidor Rabi, asked "who ordered that?" This question remains unanswered today.

The third particle in Table 1, the "pion", is very different. It was predicted to exist by a Japanese physicist, Hideki Yukawa, in 1935 (Yukawa 1935). Yukawa hypothesised that nucleons[4] in nuclei could be held together by exchanging pions between them, rather like beach-goers tossing beach-balls between them as depicted in Fig. 4. Yukawa showed that at the quantum level the process of pion exchange could give rise to an attractive force, and he was able to estimate the mass of the pion from the known range (see above) of the nuclear force. The pion was found in the cosmic radiation at the predicted mass twelve years later (Lattes et al. 1947).

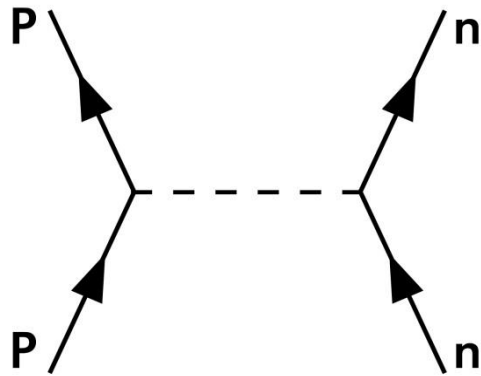

*Figure 4. Schematic diagram of Yukawa's process in which a proton and a neutron exchange a pion shown by the dashed line. Yukawa showed that the process can lead to an attractive force.*

The discovery of the pion brought on a rash of further discoveries of particles that, like the muon, were unexpected. They were aptly dubbed the "strange" particles, and the reason for their existence is as mysterious today as that for the muon (Feynman 1985). Despite this, the discoveries of the strange particles spawned the concept of the "quark" which has dominated nuclear and particle physics ever since. This is described below.

The neutrino was predicted by Pauli in the 1930s to account for the conservation of energy in β radioactivity (Pauli 1933). The neutrino interacts extremely weakly with matter and was first observed by Reines & Cowan (1956) in the intense flux from a nuclear power reactor.

The remaining particle type in the table is the "resonance". Many resonances have been discovered, and only some of the first ones to be found are listed in the table. They can be thought of as excited strongly interacting particles. Just as an atom can be excited and decay rapidly with the emission of a photon, so can a strongly interacting particle be excited and decay very rapidly with the emission of a pion. Neutrons and protons within an atomic nucleus can exchange ρ and ω resonances as well as Yukawa's pions. The processes of ρ and ω exchange are thought to provide the repulsive short range nuclear force implied by Figure 2 (Machleidt 1989).

---

[4] "Nucleon" is the generic term signifying either a proton or a neutron.

To summarize, the number particles known by the early sixties was large, of order 100 when all charge states were included. The vast majority of these particles were strongly interacting, but there were no clear patterns amongst the particle "zoo". Theoretical considerations and elementary observations from the 1930s suggested the existence of the $e^+$, π, ρ, ω and ν particles, but the muon and the strange particles were unexpected.

*Naïve quarks*

As early as 1949 Enrico Fermi and C.N. Yang had suggested there were too many known types of particles for them all to be elementary, point particles. They proposed (Fermi & Yang 1949) that Yukawa's pion was a bound state of a nucleon and an antinucleon[5].

Their idea was generalized by Gell-Mann in 1964 when he proposed that all the strongly interacting particles in Table 1 were bound states of a small set of elementary particles which he termed "quarks" (Gell-Mann 1964). In this way Gell-Mann attempted to give some order to the particle zoo. According to the model, atoms are composed of electrons and nuclei, nuclei of nucleons, and nucleons of quarks. It therefore assumed that the quark is the fundamental constituent at the heart of all matter.

Gell-Mann was able to make an impressive correspondence between the known strongly interacting particles, including those in Table 1, and the bound states of quarks following simple rules. However, the model was at best a stopgap, as it did not include forces to bind quarks together. Also, it assumed that the electric charges of quarks were fractional parts of the charge of the proton, either $+⅔$ or $-⅓$. This seemed at the time to be inelegant or surprising. For these reasons the model was received with some scepticism, including by Gell-Mann in his 1964 paper. It became to be known as the naïve quark model.

*Unified quarks*

As an MSc student at the University of Auckland in the early 1960s I studied the interactions of the strongly interacting particles listed in Table 1. This was the problem that Rutherford and Chadwick had commenced in the 1920s (Rutherford & Chadwick 1925). This eventually led, as described below, to the proposal of a speculative theory that attempted to unify the interactions of quarks with electromagnetism.

The standard formalism for conducting computations in particle physics in the 1960s was, as it is now, one known as "quantum field theory". This combines quantum theory with relativity theory in the simplest known way. It was developed by Paul Dirac, Enrico Fermi, Wolfgang Pauli, Julian Schwinger, Richard Feynman, Freeman Dyson and others in the 1930s and 1940s.

When applied to electrons, photons and positrons the formalism yields the most precise predictions yet known in science, to about ten significant figures (Feynman 1985). But there is a price. The equations need first to be "renormalized" to obtain this astounding accuracy, and the renormalization process has not been shown to be mathematically consistent. Indeed, it appears to introduce divergences that render the theory self-inconsistent (Dirac 1958; Feynman 1985).

I first learned of the renormalization problem while introducing myself to quantum field theory during my MSc studies. Despite the astounding successes of the renormalisation theory, I was unable to convince myself of its validity. Later, while at MIT in the early 1960s as a PhD student, I learned of work underway there on the renormalization problem by a group led by one of their then-young theorists, Kenneth Johnson. I assisted with some calculations of this group (Rosner 1966) whilst thinking about possible implications.

---

[5] "Antinucleon" is the generic term signifying either an antiproton or an antineutron.

Late in the 1960s I returned to New Zealand after spells in the US, Italy and Switzerland and proposed a scheme whereby it seemed that it might be possible to solve the problems of renormalization theory and naïve quarks simultaneously (Yock 1969). This entailed assuming that quarks carry dual charges, where the first charge was normal (non-fractional) electric charge and the second a high or strong charge. The strong charge appeared to offer a possible means for solving the renormalization problem, and also a mechanism for binding quarks within nucleons via the strong attraction that would occur between quarks with strong positive charges and antiquarks with strong negative charges.

I further assumed that nucleons would combine to form nuclei much as atoms combine to form molecules (Yock 1970; Anon. 1971). In other words, the force that binds quarks together in nucleons was assumed also to be the source of the strong nuclear force that Rutherford and Chadwick commenced studying in the 1920s. The theory was quite economical in this sense. Furthermore, it provided possible reasons for the existence of particles such as the muon and the strange particles. However, the correspondence between predicted and known particles was qualitative at best.

The reasoning underlying the model entailed many assumptions that are still under investigation today, in particular the properties of quantum field theory when high charges are present (Kizilersü et al. 2013). The fundamental assumption was the neutrality of bound states of quarks with respect to the strong charge. I expressed this with the following words *"hence it is quite plausible, and this is a basic postulate, that all readily observable states are hadronically neutral"* (Yock 1969).

***Coloured quarks***
Four years later a theory of coloured quarks was proposed at Caltech (Fritzsch et al. 1973). This now forms the basis of today's so-called Standard Model of matter (Anon. 2016a). It includes some of the above concepts in modified form as described below.

Fritzsch et al. assumed, as in the unified theory above, that quarks carry dual charges where one was electric and the other a strong charge. But they assumed the strong charge was not electric, instead they assumed it to be a new type of charge which they termed "colour" charge. Instead of being merely positive or negative, colour charge was assumed to occur in three varieties which were termed "red", "green" and "blue". Neutrality with respect to colour charge was assumed to be achieved by either combining quarks with antiquarks or by combining equal mixtures of quarks of the three colours.

Fritzsch et al expressed the assumption of colour neutrality with the words *"then it is easy to envisage a situation in which the only states with deep attraction would be colour singlets"* that were startlingly reminiscent of those I had used in 1969. The paper by Fritzsch et al. escaped my attention until it was reviewed in the 1990s at which time it struck me that my earlier ideas on dually charged quarks and the neutrality of bound states may have had some impact.

Some 40 years have elapsed since the above assumptions of the neutrality of bound states of quarks were made, but neither has been proven. This is not for want of trying. A US$1M prize has been on offer since 2000 by the Clay Mathematics Institute of New Hampshire for what amounts to a proof of the assumption of colour neutrality, but the prize remains unclaimed (Anon. 2016b).

Meanwhile the assumption of colour neutrality is taken for granted by virtually all workers in particle physics today. It is assumed that the colour binding mechanism is so strong that quarks are permanently confined to the interiors of particles such as protons and neutrons. It is also assumed, as I had done in 1969, that neighbouring neutrons and protons in atomic nuclei exert forces upon one another in much the same way as atoms do in molecules. This is now textbook material (Anon. 2016a).

Questions remain however. The colour theory assumes, for example, that quarks bind in pairs and triplets only, but not in larger combinations. In recent years evidence has been found for several four-quark and five-quark combinations (Cho 2016). Also, the colour binding mechanism is not explicitly included in calculations. It is merely assumed to occur.

Historical questions also arise. It is claimed, for example, that the strong nuclear force does not arise through Yukawa's mechanism of pion exchange supplemented by ρ and ω exchange (Ishi et al. 2007). The successful prediction of the mass of the pion by Yukawa, and also the subsequent discoveries of the ρ and ω particles, appear to arise as fortuitous accidents according to the colour theory, as do several other successful calculations involving these particles at low energies (Machleidt 1989; Vanderhaeghen & Walcher 2011), at medium energies (Yock & Gordan 1967; Yock 1968) and at high energies (Boros & Zuo-tang 1995; Derrick et al. 1996; Thomas & Boros 1999; D'Alesio et al. 2000; Lu et al. 2000). The nature of these apparent accidents requires further understanding.

More dramatically, the existence of the strange particles in Table 1 is not explained by the model. This is particularly vexing, as it was the discovery of these particles which more than anything else led to the introduction of the quark idea. It is a trivial matter to excise the strange and related particles, such as the muon, from the Standard Model. The result is a mathematically consistent theory of undeniably greater simplicity and elegance than the Standard Model. However, it would not apply to the real world. So Rabi's question from 80 years ago is not answered by the Standard Model. Indeed, it has become more pressing. Not only does the muon require explanation, the strange particles and many more particles also do.

The problems of renormalization theory have also not been solved by the model. And the discovery of the Higgs boson at the Large Hadron Collider that was reported in 2012 may yet prove to be problematical. Results from the Large Hadron Collider from 2015 indicated the possible presence, as yet unpublished, of a similar but unexpected particle. This perplexing situation will be studied further in the future.

Other problems could be cited. The author's personal opinion is that the Standard Model raises significant problems and that, instead of tinkering with it, a fresh start may well be needed. I see this as an exciting prospect even though it is a minority point of view at the present time.

### *The status of the 1969 theory at the University of Auckland*
As described above, today's Standard Model posits that the nucleons of the atomic nucleus are composed of dually charged quarks which possess electric and colour charges. This was proposed in 1973. The colour charges are assumed to neutralize one another in nucleons and they are assumed to be the source of the strong nuclear force that Rutherford and Chadwick first explored in 1925. The latter force is assumed in the Standard Model to be comparable to the chemical bonding that acts between atoms in molecules.

As was also noted above, the concepts of dually charged quarks, neutrality of bound states, and chemical-like interactions in nuclei were proposed in New Zealand in 1969 in a theory that may be regarded as a precursor to the colour theory. These concepts were new at the time and I believe that New Zealand's isolation from the large northern hemisphere centres of learning aided their formulation. Despite our isolation the concepts were noted at the time (e.g., Anon. 1971). Also, as noted above, the language they were proposed in mirrored that used in the subsequent development of the Standard Model. They may therefore have been influential.

Unfortunately, present and former colleagues at the University of Auckland who question the work described above on planets also question the relationship between the Standard Model and the above work carried out in New Zealand.

### *Newton's hypothesis of the conformability of nature*
In 1704 Newton published far-reaching speculations on a number of topics in the closing pages of his treatise on *"Opticks"* (Newton 1704). In particular he hypothesized that nature would be *"conformable to herself"* at different scales, or, in modern terminology, "self-similar". He stated this hypothesis more than once, for example with the following words:-

> *"There are therefore Agents in Nature able to make the Particles of Bodies stick together by very strong Attractions. And it is the Business of experimental Philosophy to find them out. Now the smallest Particles of Matter may cohere by the strongest Attractions, and compose bigger Particles of weaker Virtue; and many of these may cohere and compose bigger Particles whose Virtue is still weaker, and so on for divers Successions, until the Progression end in the biggest Particles on which the Operations in Chymistry, and the Colours of natural Bodies depend, and which by cohering compose Bodies of a sensible Magnitude."*

These words seemed eminently reasonable to the author when the above ideas on dually charged quarks and the neutrality of bound states were proposed, and Newton's words were reproduced in a publication at the time (Yock 1970). Acording to the unified quark model the binding within atoms and nucleons are both consequences of neutrality principles, and the binding between atoms and nucleons are both consequences of chemical or chemical-like principles. Self-similarity thus occurs twice over scales that differ greatly.

Others subsequently reproduced Newton's words, in particular Gell-Mann in 2009 and Steven Weinberg in 2015 (Gell-Mann 2009; Weinberg 2015). Both were leading contributors to the development of the Standard Model and it is interesting to ask whether or not the Standard Model satisfies the principle of self-similarity. One can argue that it introduces new concepts at the deepest level of matter (confined, coloured and fractionally charged quarks) and that therefore it is not self-similar, but one can also argue that it follows a gauge principle[6] at all scales, so the question is moot. I believe it is fair to say that the unified theory of quarks (also a gauge theory) satisfies the principle of self-similarity from the scale of molecules to that of quarks.

## Discussion and conclusions
The Oxford Dictionary defines the Scientific Method as a method of procedure that has characterized natural sciences since the 17th century, consisting in systematic observation, measurement, and experiment, and the formulation, testing, and modification of hypotheses (Anon. 2016c).

If the method had been followed in the research described above, especially the call for systematic observations of planets, then the deleterious actions described above could not have occurred.

What could be less systematic than disregarding confirmed results, or using computer code with known errors? And what could be more systematic than Newton's hypothesis of self-similarity?

---

[6] The gauge principle (Weyl 1928) in quantum field theory is a generalization of the principle one learns in physics at high-school that the voltage at any point in an electrical circuit is arbitrary, only voltage differences between pairs of points are defined. Quantum field theories that maintain the gauge principle incur fewer divergences in the renormalization process than those that do not.

It is of course perfectly acceptable to re-analyse data in any research project. Mistakes can be made. A recent study conducted at Cambridge University on ancient human genome that had been retrieved from the skeleton of a man who lived in Ethiopia 4,500 years and which appeared to have unexpected implications turned out to be flawed by human error (Zimmer 2016).

However, in the case of the fourth microlensing planet described above, there was no reason to question the original analyses from 2006 as consistent results had been obtained independently in the US and in New Zealand. Turning a blind eye to the NZ results that had been obtained using normal procedures did not negate them. Likewise, turning a blind eye to the recent results from the Hubble Space Telescope and the Keck Observatory did not negate them.

Similarly, to turn a blind eye to similarities between published ideas from New Zealand on the nature of the strong nuclear force and ideas that subsequently surfaced in today's Standard Model of matter seems, to the author, to indicate a dismal viewpoint of the scientific potential of small countries.

Many examples could of course be recounted, in science and other endeavours, to remind ourselves that New Zealand's smallness and isolation need not be barriers to originality or success. The late Dan Walls, a highly successful pioneer in the field of quantum optics and a keen sportsman, was proud to work in New Zealand, and the Geothermal Institute at the University of Auckland has made excellent use of one of our natural resources and attracted international students for many years. Antarctica awaits the NZ astronomical community as possibly providing the best sites on the surface of our planet, such as Ridge A at an altitude of 4000m and latitude $80°S$, for astronomy (Freeman 2016). Planets orbiting nearby stars could be found from such a site and subsequently examined for the presence of bio-signatures in their atmospheres (Yock 2016; Hecht 2016).

I believe our scientific community should build on the special opportunities our location on the globe offers, not the reverse. As Dame Anne says, we should be inventive, entrepreneurial, exciting and generous in spirit (Salmond 2015).

## Acknowledgement

I thank the Editorial Committee for helpful comments on the manuscript.

*****************************************